\documentclass[useAMS,usenatbib]{mn2e}

\usepackage[dvips]{graphicx}
\usepackage{amssymb}
\usepackage{amsmath}
\usepackage{amstext}

\newcommand{\adfo}{${\rm ADF(O^{2+})}$}

\newcommand{\te}{$T_{e}$}
\newcommand{\nel}{$n_{e}$}

\newcommand{\tf}{$t^{2}$}
\newcommand\ion[2]{#1~{\sc {#2}}\relax}        
\newcommand\ioni[2]{${\rm #1^{#2}}$}           

\newcommand{\cmc}{{\rm cm$^{-3}$}}
\newcommand{\kms}{{\rm km~s$^{-1}$}}

\newcommand{\hh}{HH~202}
\newcommand{\hii}{H~{\sc ii}}
\newcommand{\hi}{H~{\sc i}}

\title[Integral field spectroscopy of \hh]{Properties of the ionized gas in \hh. I: Results from 
                                           integral field spectroscopy with PMAS%
			       \thanks{Based on observations collected at the Centro Astron\'omico 
			       Hispano Alem\'an (CAHA) at Calar Alto, operated jointly by the 
			       Max-Planck Institut f\"ur Astronomie and the Instituto de 
			       Astrof{\'{\i}}sica de Andaluc{\'{\i}}a (CSIC).}}
\author[A. Mesa-Delgado et al.]
       {A. Mesa-Delgado$^1$\thanks{E-mail: amd@iac.es}, 
        L. L\'opez-Mart{\'{\i}}n$^1$, C. Esteban$^1$, J. Garc{\'{\i}}a-Rojas$^2$, 
	\newauthor 
	and V. Luridiana$^3$\\
	$^1$Instituto de Astrof\'\i sica de Canarias, E-38200 La Laguna, Tenerife, Spain \\
        $^2$Instituto de Astronom\'\i a, UNAM, Apdo. Postal 70-264, 04510 M\'exico D.F., Mexico\\
        $^3$Instituto de Astrof\'\i sica de Andaluc\'\i a (CSIC), Apdo. Correos 3004, E-18080 
	    Granada, Spain\\}

\begin{document}

\date{Accepted X XX XX. Received XXXX XX XX; in original form XXXX XX XX}
\pagerange{\pageref{firstpage}--\pageref{lastpage}} \pubyear{2008}

\maketitle
\label{firstpage}

\begin{abstract}
 We present results from integral field spectroscopy with the Potsdam Multi-Aperture Spectrograph 
 of the head of the Herbig-Haro object \hh\ with a spatial sampling of 1$\arcsec\times$1$\arcsec$. 
 We have obtained maps of different emission lines, physical conditions --such as electron 
 temperature and density-- and ionic abundances from recombination and collisionally excited lines. 
 We present the first map of the Balmer temperature and of the temperature fluctuation parameter, 
 \tf. We have calculated the \tf\ in the plane of the sky, which is substantially smaller than that 
 determined along the line of sight. We have mapped the abundance discrepancy factor of 
 \ioni{O}{2+}, \adfo, finding its maximum value at the \hh-S position. We have explored the 
 relations between the \adfo\ and the electron density, the Balmer and [\ion{O}{iii}] temperatures, 
 the ionization degree as well as the \tf\ parameter. We do not find clear correlations between 
 these properties and the results seem to support that the ADF and \tf\ are independent phenomena. 
 We have found a weak negative correlation between the \ioni{O}{2+} abundance determined from 
 recombination lines and the temperature, which is the expected behaviour in an ionized nebula, 
 hence it seems that there is not evidence for the presence of super-metal rich droplets in \hii\ 
 regions.  
\end{abstract}


\section{Introduction} \label{intro}
\begin{figure*}
   \centering
   \includegraphics[scale=0.7]{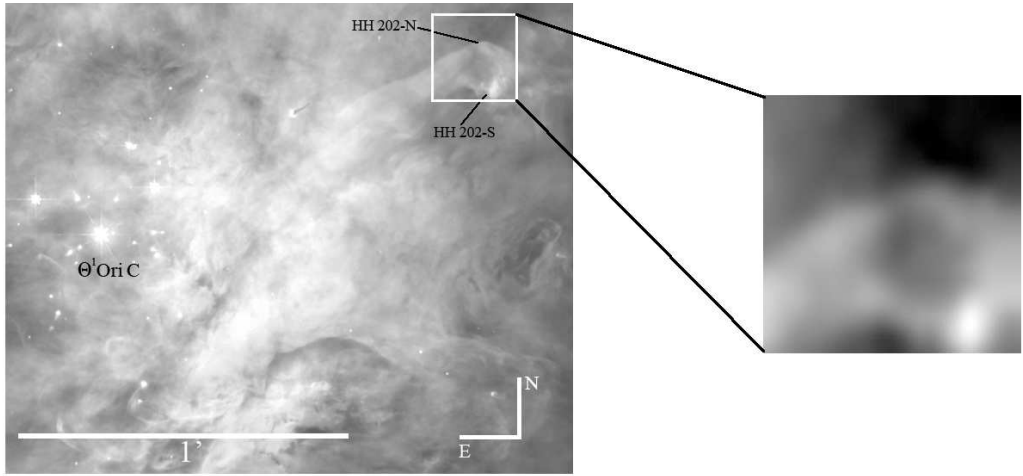} 
   \caption{$HST$ image of the central part of the Orion Nebula, which combines WFPC2 images taken 
            in different filters \citep{odellwong96}. The white square corresponds to the field of 
	    view of PMAS IFU used, covering the head of \hh. The separate close-up image on the 
	    right shows the H$\alpha$ map obtained with PMAS. The original map is 16$\times$16 
	    pixels of 1$\arcsec\times$1$\arcsec$ size and has been rebinned to 160$\times$160 
	    pixels. Note the remarkable similarity between the $HST$ image and our rebinned PMAS 
	    H$\alpha$ map.}
   \label{f1}
  \end{figure*}

 The Orion Nebula is the nearest, most observed and studied Galactic \hii\ region. It is an 
 active star-formation region where phenomena associated with the early stages of stellar evolution 
 such as protoplanetary disks (proplyds) and Herbig-Haro (HH) objects can be observed in detail.\\
 HH objects are bright nebulosities associated with high-velocity gas flows. The most prominent 
 high-velocity feature in the nebula is the Becklin-Neugebauer/Kleinmann-Low (BN/KL) complex, which 
 contains several HH objects (HH~201, 205, 206, 207 and 208). But also, there are other important 
 high-velocity flows that do not belong to the BN/KL complex, as is the case of \hh, 203 and 204. 
 The origin of these flows has been associated with infrared sources embedded within the Orion 
 Molecular Cloud 1 South \citep[see][and references therein]{odellhenney08}.\\
 \hh\ was one of the first HH objects identified in the Orion Nebula by \cite{cantoetal80} in 
 their [\ion{S}{ii}] and [\ion{N}{ii}] images as an emission line object showing two bright knots 
 --the so-called \hh-N and \hh-S (see Figure \ref{f1})-- embedded in a more extended nebulosity 
 with a long concave form. After its discovery, the spectroscopic work by \cite{meaburn86} showed 
 that the arc-shaped nebulosity emits in [\ion{O}{iii}] lines.\\
 Further high-spectral resolution spectroscopy in several emission lines ([\ion{O}{i}], 
 [\ion{S}{ii}], [\ion{N}{ii}], [\ion{O}{ii}], [\ion{S}{iii}] and H$\alpha$) has been performed in 
 \hh\ by \cite{odelletal91} who studied \hh-S and found two velocity components. More recently, we 
 can find in the literature extensive works on the gas kinematics in the Orion Nebula as those by 
 \cite{doietal02} and \cite{odelldoi03}, where tangential velocities have been measured, or 
 \cite{doietal04}, who provide with accurate radial velocity measurements, detecting up to three 
 kinematic components in \hh-S. On the other hand, the most detailed imaging study of the Orion 
 Nebula in the [\ion{S}{ii}], [\ion{O}{iii}] and H$\alpha$ lines has been carried out by 
 \cite{odelletal97} with the Hubble Space Telescope ($HST$). These authors also detect an intense 
 [\ion{O}{iii}] emission in the extended nebulosity of \hh\ and a strong [\ion{S}{ii}] emission at 
 its knots. The extended [\ion{O}{iii}] emission found in all the works indicates that the main 
 excitation mechanism of \hh\ is photoionization, most probably from the brightest and hottest star 
 of the Trapezium cluster, $\theta^1$ Ori C. This particular property --not very common in other 
 HH objects, which are usually ionized by shocks-- permits to derive the physical conditions and 
 chemical abundances of the high velocity gas associated with \hh\ making use of the standard 
 techniques for the study of ionized nebulae.\\
 Our group has studied in detail the chemical composition of the Orion Nebula in some previous 
 papers \citep{estebanetal98,estebanetal04,mesadelgadoetal08} and is especially interested in 
 studying the so-called abundance discrepancy (AD) problem in \hii\ regions, which is the 
 disagreement between the abundances of the same ion derived from collisionally excited lines 
 (CELs) and recombination lines (RLs). From intermediate and high resolution spectroscopy of 
 Galactic and extragalactic \hii\ regions, our group has found that the \ioni{O}{2+}/\ioni{H}{+} 
 ratio calculated from RLs is always between 0.1 and 0.3 dex higher than the value obtained from 
 CELs for the same region \citep[see][]{garciarojasesteban07}. The results obtained for \hii\ 
 regions are quite different to those obtained for planetary nebulae (PNe), where the AD shows a 
 much wider range of values and could be substantially larger in some objects \citep[$e.g.$][]%
 {liuetal00,liuetal06,tsamisetal03b,tsamisetal04}. As argued in \cite{garciarojasesteban07}, the 
 AD problem in \hii\ regions seems to be consistent with the predictions of the temperature 
 fluctuaction paradigm proposed by \cite{peimbert67} and parametrized by the mean square of the 
 spatial distributions of temperature, the so-called \tf\ parameter. In this scenario, the AD 
 would be produced by the very different temperature dependence of the emissivities of both kinds 
 of lines. In any case, the existence and origin of such temperature fluctuations are still 
 controversial. \cite{tsamispequignot05} and \cite{stasinskaetal07} have proposed a different 
 hypothesis in order to explain the AD in \hii\ regions. Based on the model for heavy-element 
 mixing of \cite{tenoriotagle96}, the presence of cold metal-rich droplets of supernova ejecta 
 still not mixed with the ambient gas of the \hii\ regions would produce most of the RL emission 
 while CEL emission would be produced by the ambient gas, wich would have the typical electron 
 temperature of an \hii\ region and the expected composition of the local interstellar medium. In 
 such case, the abundances from CELs would be the truly representative ones of the nebula.\\
 Several authors have studied the spatial distribution of the physical conditions in the Orion 
 Nebula. \cite{odelletal03} have obtained a high-spatial resolution map of the electron temperature 
 in a field centered at the southwest of the Trapezium cluster from narrowband images taken with 
 the $HST$. They found small spatial scale temperature variations, which seem to be in quantitative 
 agreement with the \tf\ parameter determined using the AD factor found by \cite{estebanetal04}. On 
 the other hand, \cite{rubinetal03} have taken long-slit spectroscopy making use of $STIS$ at the 
 $HST$ in several positions and do not find substantial spatial variations or gradients of the 
 physical conditions along their slits positions. More recently, \cite{sanchezetal07} have obtained 
 a mosaic of the Orion Nebula from integral field spectroscopy with a spatial resolution of 
 2\farcs7, but lacking a confident absolute flux calibration due to bad transparency during the 
 observations. Their electron temperature map shows clear spatial variations and the electron 
 density map is very rich in substructures, some of which could be related to HH objects. A more 
 detailed view --but based on long-slit spectroscopy-- has been obtained by 
 \cite{mesadelgadoetal08}, who obtained spatial distributions --at spatial scales of 1\farcs2-- of 
 electron temperature, density, line intensities, ionic abundances and the AD factor (ADF, defined 
 as the difference between the abundances derived from RLs and CELs) for five slit positions 
 crossing different morphological zones of the Orion Nebula, finding spikes in the distribution of 
 the electron temperature and density which are related to the position of proplyds and HH objects. 
 In particular, these authors found that the ADF shows larger values at the HH objects.\\
 The use of the integral field spectroscopy is still in its infancy in the study of ionized nebulae 
 and the number of works available is still rather small. In the case of PNe, \cite{tsamisetal08} 
 have carried out the first deep study using this technique with FLAMES at the VLT taking spectra 
 of three PNe of the Galactic disc and covering the spectral range from 3964 to 5078 \AA. For 
 the Orion Nebula, \cite{tsamisetal08b} have presented preliminary results of integral field 
 spectrocopy with FLAMES for several proplyds with the aim of studying the behaviour of the 
 physical conditions, chemical abundances and the ADF in and outside these kinds of objects. The 
 first results show that the temperature measured in the proplyds can be affected by collisional 
 deexcitation as has been also suggested in previous works \citep{rubinetal03,%
 mesadelgadoetal08}.\\
 The main goal of the present paper is to use integral field spectroscopy at spatial scales of 
 1$"$ in order to explore the effect of HH objects on the derived ADF considering the results of 
 \cite{mesadelgadoetal08}. \\ 
 In \S\ref{obsred} we describe the observations obtained with PMAS and the reduction procedure. In 
 \S\ref{medpmas} we describe the emission line measurements and the reddening correction as well 
 as some representative bidimensional maps of those quantities. In \S\ref{resul} we describe the 
 determination of the physical conditions, chemical abundances --from both kinds of lines, CELs 
 and RLs-- and the ADF of \ioni{O}{2+} as well as their corresponding maps. In \S\ref{discu} we 
 calculate the \tf\ parameter from different methods and present the first bidimensional map of 
 this quantity. Also, we show the possible correlations among the ADF and the rest of properties 
 determined. Finally, in \S\ref{conclu} we summarize our main conclusions. 
\section{Observations and Data Reduction} \label{obsred}
 {\hh} was observed on 2007 October 14 at Calar Alto Observatory (Almer\'ia, Spain), using the 
 3.5m Telescope with the Potsdam Multi-Aperture Spectrometer \citep[PMAS,][]{rothetal2005}. The 
 standard lens array integral field unit (IFU) of 16$\arcsec\times$16$\arcsec$ field of view (FOV) 
 was used with a sampling of 1$\arcsec$. Most of the optical range was covered with the V600 
 grating using two grating rotator angles: $-$72, covering from 3500 to 5100 \AA; and $-$68, 
 covering from 5700 to 7200 \AA. The effective spectral resolution was 3.6 \AA. The position of 
 the IFU covering \hh\ is shown in Figure~\ref{f1} where we can see the two bright knots \hh-N and 
 \hh-S together with an H$\alpha$ image obtained with PMAS rebinned to 160$\times$160 pixels using 
 the {\sc rebin} function of {\sc idl} which performs an expansion of the original map through a 
 linear interpolation. The blue and red spectra have a total integration time of 1800 and 1000 
 seconds, respectively. Additional short exposures of 10 seconds were taken in order to avoid 
 saturation of the brightest emission lines. Calibration images were obtained during the night: arc 
 lamps for the wavelength calibration and a continuum lamp needed to extract the 256 individual 
 spectra on the CCD. Observations of the spectrophotometric standard stars BD~$+$28$\rm ^o$4211, 
 Feige 110 and G~191-B2B \citep{oke90} were used for flux calibration. The error of this 
 calibration is of the order of 5\%. The night was photometric and the typical seeing during the 
 observations was 1$\arcsec$.\\ 
 The data were reduced using the {\sc iraf} reduction package {\sc specred}. After bias 
 subtraction, spectra were traced on the continuum lamp exposure obtained before each science 
 exposure, and wavelength calibrated using a HgNe arc lamp. The continuum lamp and sky flats were 
 used to determine the response of the instrument for each fiber and wavelength. Finally, for the 
 standard stars we have co-added the spectra of the central fibers and compared them with the 
 tabulated one-dimensional spectra.\\
 We have noticed the effect of the differential atmospheric refraction (DAR) in the monochromatic 
 images of \hh\ obtained for Balmer lines at different wavelengths reaching the value of $\sim$%
 1\farcs3 between H$\alpha$ and H11. We have measured offsets between all Balmer line images, 
 and shifted with respect to H$\alpha$. The maximum coincident FOV resulting in the whole 
 wavelength range is 15$\arcsec\times$15$\arcsec$. All the maps involving emission line ratios 
 analysed in this paper have been corrected for DAR.\\
 Due to the relatively low spectral resolution of the observations, Hg 4358 \AA\ telluric emission 
 was somewhat blended with [\ion{O}{iii}] 4363 \AA\ and a proper sky subtraction was necessary. We 
 have taken a median value of the Hg 4358 \AA\ line flux (9.22 $\times 10^{-16}$ 
 erg~cm$^{-2}$~s$^{-1}$), which is in some regions approximately two orders of magnitude lower than 
  \begin{figure}
   \centering
   \includegraphics[scale=0.45]{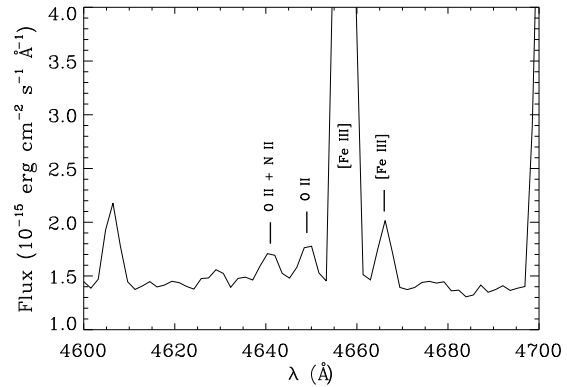} 
   \caption{Section of a PMAS spectra around \ion{O}{ii} lines of multiplet 1 corresponding to 
            the spaxel position (-5.5,~-6.5).} 
   \label{espmas}
  \end{figure}    
 the auroral line flux.\\  
 In previous works the background nebular emission in Orion proplyds has been determined, either 
 using echelle data \citep{henneyodell99} or integral field spectroscopy \citep{vasconcelosetal05}. 
 In our case, the low spectral resolution of the data does not allow to separate the \hh\ emission 
 from the background. Furthermore, as pointed out in both cited papers, this task is very 
 challenging due to small-scale inhomogeneities in the nebular emission. As a result, only rough 
 estimates can be given. Using the method proposed by \cite{vasconcelosetal05}, we have assumed 
 that the background emission is represented by the emission in the northeast corner of the FOV. 
 The resulting background contribution is thus estimated to lie between 40\% and 50\% in knot 
 \hh-S, and between 55\% and 65\% in the rest of \hh. If such a correction were applied to our 
 data, we would obtain much larger densities, lower temperatures from the [\ion{N}{ii}] line ratio 
 and higher temperatures from the [\ion{O}{iii}] line ratio with respect to the results presented 
 in next sections. These results do not change the final conclusions of the paper because we are 
 interested in how the presence of an HH object affects to the ADF when low spectral resolution 
 observations are made and the contributions from the gas flows and the nebular background are 
 mixed. In Paper II, we perform a complete analysis of the knot \hh-S based on echelle 
 spectrophotometry. The high resolution of these data has allowed us to resolve the kinematic 
 component associated with the gas flow and the one associated with the background emission (in a 
 small area included in the PMAS field). This analysis indicates that the background emission 
 ranges from 30\% in the case of [\ion{N}{ii}] 5755 \AA\ to 60\% for [\ion{O}{iii}] 5007 \AA. In 
 the case of \hi\ lines, the background emission is about 50\%.
  \begin{figure*} 
   \centering
   \includegraphics[scale=0.6]{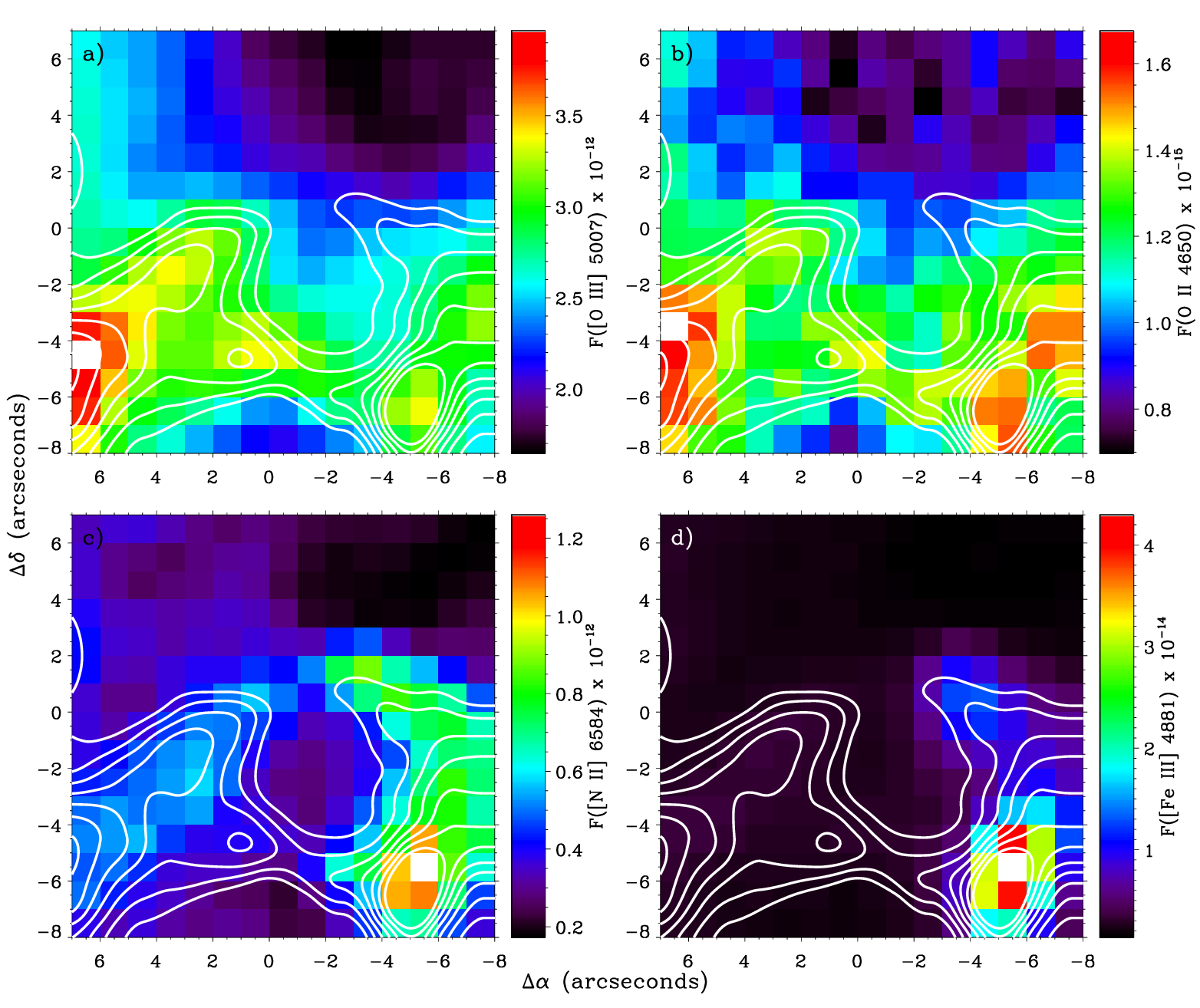} 
   \caption{Emission line flux maps with H$\alpha$ contours overplotted on the \hh\ PMAS field
            of: a) [\ion{O}{iii}] 5007 \AA, b) \ion{O}{ii} 4650 \AA, c) [\ion{N}{ii}] 6584 
            \AA\  and d) [\ion{Fe}{iii}] 4881 \AA\ (in units of erg~cm$^{-2}$~s$^{-1}$).}
   \label{flux}
  \end{figure*}	
\section{Line measurements and reddening correction} \label{medpmas}
\begin{figure*}
    \centering
    \includegraphics[scale=0.6]{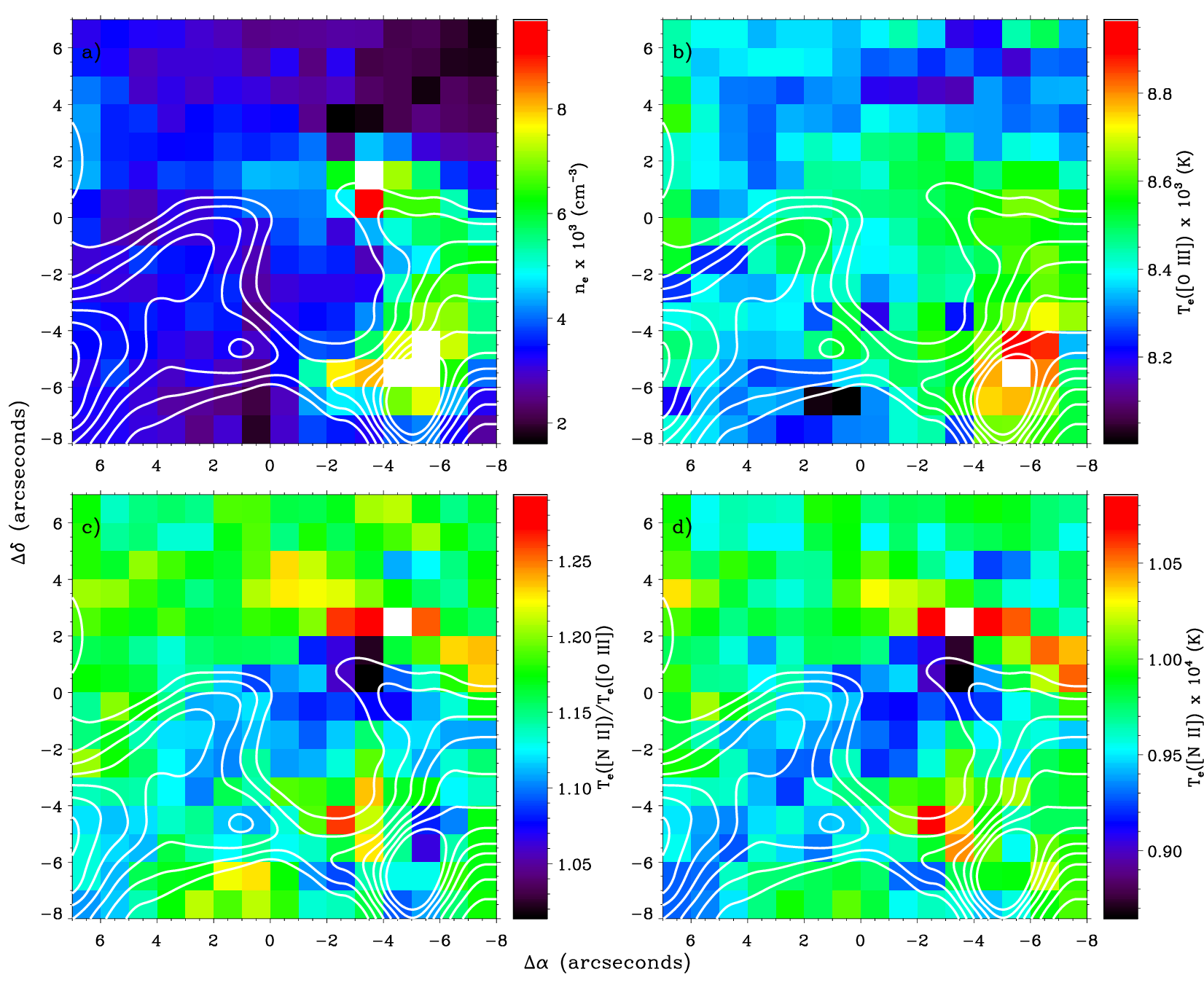} 
    \caption{Physical conditions maps with H$\alpha$ contours overplotted: a) \nel([\ion{S}{ii}]),
             b) \te([\ion{O}{iii}]), c) \te([\ion{N}{ii}])/\te([\ion{O}{iii}]) ratio and 
	     d) \te([\ion{N}{ii}]).}
    \label{condfis}
   \end{figure*}

 The emission lines considered in our analysis are the following: a) Balmer lines, from H$\alpha$ 
 to H11, which are used to compute the reddening correction and correct the DAR shift (H8 and 
 H10 were not used because they suffer from line-blending); b) CELs of various species, which are 
 used to compute physical conditions and ionic abundances; c) Faint RLs of \ion{C}{ii} and 
 \ion{O}{ii} which are used to derive the ionic abundances and to compute the \adfo. In Figure 
 \ref{espmas}, we show an example of the spectra of a given spaxel where we can see the blend of  
 \ion{O}{ii} lines at 4649 and 4651 \AA.\\
 Line fluxes were measured applying a single or a multiple Gaussian profile fit procedure between 
 two given limits and over the local continuum. All these measurements were made with the 
 {\sc splot} routine of {\sc iraf} and using our own scripts to automatize the process. The errors 
 associated with the line flux measurements were determined following \cite{mesadelgadoetal08}. The 
 final error of a line was computed as the quadratic sum of the error in its flux measurement and 
 the error in flux calibration. In order to avoid spurious weak line measurements, we imposed three 
 criteria to discriminate between real features and noise: 1) Line intensity peak over 2.5 times 
 the sigma of the continuum; 2) FWHM(\ion{H}{i})/1.5 $<$ FWHM($\lambda$) $<$ 
 1.5$\times$FWHM(\ion{H}{i}); and 3) F($\lambda$) $>$ 0.0001 $\times$ F(H$\beta$).\\
 All line fluxes of a given spectrum have been normalized to H$\beta$ and H$\alpha$ for the blue 
 and red range, respectively. To produce a final homogeneous set of line ratios, all of them were 
 re-scaled to H$\beta$. The re-scaling factor used in the red spectra was the theoretical 
 H$\alpha$/H$\beta$ ratio for the physical conditions of \te = 10000 K and \nel = 1000 \cmc.\\
 In Figure \ref{flux}, we show [\ion{O}{iii}] 5007 \AA, \ion{O}{ii} 4650 \AA, [\ion{N}{ii}] 6584 
 \AA\ and [\ion{Fe}{iii}] 4881 \AA\ emission line maps with H$\alpha$ contours overplotted. 
  \begin{figure*}
   \centering
   \includegraphics[scale=0.65]{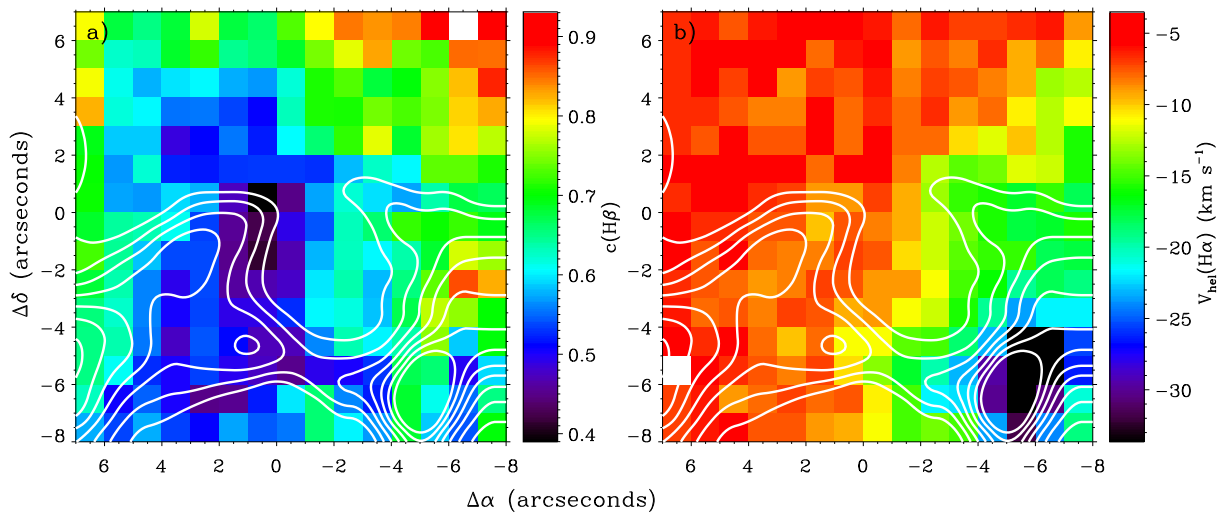} 
   \caption{\hh\ PMAS maps with H$\alpha$ contours overplotted. a) c(H$\beta$),  
	     b) H$\alpha$ heliocentric velocity.}
   \label{chb}
  \end{figure*}	
 In Figures~\ref{flux}a and \ref{flux}b, it is possible to see that the [\ion{O}{iii}] 5007 and 
 \ion{O}{ii} 4650 \AA\ spatial distributions are quite similar. In particular, the \ion{O}{ii} flux 
 map seems to peak at the position of the apex of \hh, the zone also known as \hh-S --spaxels 
 (-5.5,-6.5) and (-6.5,-4.5)--. A similar spatial distribution of both kinds of lines was 
 already observed using long-slit spectra in the Orion Nebula by \cite{mesadelgadoetal08}, as well 
 as in the three PNe observed by \cite{tsamisetal08}. Figure~\ref{flux} also shows that the spatial 
 distributions of [\ion{N}{ii}] and [\ion{O}{iii}] are rather different. The [\ion{N}{ii}] map 
 shows an enhancement in the bow shock of \hh. HH objects are characterized by their strong 
 emission in [\ion{Fe}{iii}] lines. In Figure~\ref{flux}d  we can see the spatial distribution of 
 [\ion{Fe}{iii}] 4881 \AA\ where the maximum emission is strongly concentrated at \hh-S, which 
 coincides with the maxima in radial velocity and electron density as we can see in Figures 
 \ref{chb}b and \ref{condfis}a. This high [\ion{Fe}{iii}] emission at \hh-S may be related to 
 shock destruction of dust grains or to the particular excitation conditions of the gas in the 
 knot.\\
 The reddening coefficient, c(H$\beta$), has been obtained by fitting the observed 
 H$\gamma$/H$\beta$, H$\delta$/H$\beta$, H9/H$\beta$ and H11/H$\beta$ ratios to the theoretical 
 ones predicted by \cite{storeyhummer95} for \nel\ = 1000 \cmc\ and \te\ = 10000 K. We have used 
 the reddening function, $f(\lambda)$, normalized to H$\beta$ determined by \cite{blagraveetal07} 
 for the Orion Nebula. The use of this extinction law instead of the classical one of 
 \cite{costeropeimbert70} produces c(H$\beta$) values about 0.1 dex higher and dereddened fluxes 
 with respect to H$\beta$ about 3\% lower for lines in the range 5000 to 7500 \AA, 4\% higher for 
 wavelengths below than 5000 \AA\ and 2\% for wavelengths above 7500 \AA. The final adopted 
 c(H$\beta$) value for each spaxel is an average of the individual values derived from each Balmer 
 line ratio weighted by their corresponding uncertainties. The typical error of c(H$\beta$) is 
 about 0.14 dex for each spaxel. The resulting extinction map is shown in Figure~\ref{chb}a. The 
 extinction coefficient varies approximately from 0.4 to 0.7 dex and reaches the highest values 
 --between 0.8 and 0.9 dex-- in a low surface brightness area at the north of \hh. Excluding this 
 zone --which shows the largest line intensity uncertainties-- we obtain a mean value 0.56$\pm$0.14 
 dex.\\
 We have compared our c(H$\beta$) values with those obtained by \cite{odellyusefzadeh00} and 
 \cite{mesadelgadoetal08} for the area of \hh. The first authors have obtained consistent 
 c(H$\beta$) values from the H$\alpha$/H$\beta$ line ratio using calibrated $HST$ images and from 
 radio to optical surface brightness ratio. \hh\ is located between the contour lines of 0.2 and 
 0.4 dex on the maps of \cite{odellyusefzadeh00}. On the other hand, \cite{mesadelgadoetal08}, from 
 long-slit spectroscopy passing precisely through \hh-S, derive a c(H$\beta$) value of about 
 0.4$\pm$0.1 dex. Therefore, we find that our mean value of the reddening coefficient is consistent 
 with the previous determinations taking into account the errors and the different extinction laws 
 used.\\
 The heliocentric velocity map obtained from the centroid of the H$\alpha$ line profile is shown 
 in Figure~\ref{chb}b. We observe that the gas around the head of \hh\ shows a velocity of about 
 $-$5 \kms\ and that the most negatives values are reached at \hh-S position, where velocities of 
 $-$35 \kms\ are measured. These values of the velocity of \hh-S agree within the errors with that 
 determined by \cite{doietal04} of $-$39$\pm$2 \kms\ for this position. However, due to the lower 
 spectral resolution of our data, the velocities we measure correspond to the blend of the emission 
 of the background nebular gas and the gas flow of the HH object. 
\section{Results} \label{resul}
 \subsection{Physical conditions} \label{condpmas}       
  Nebular electron densities, \nel, and temperatures, \te, have been derived from the usual CEL 
  ratios --[\ion{S}{ii}] 6717$/$6731 for \nel, and [\ion{O}{iii}] (4959$+$5007)$/$4363 and 
  [\ion{N}{ii}] (6548$+$6584)$/$5755 for \te, and using the {\sc temden} task of the {\sc nebular} 
  package of {\sc iraf} \citep{shawdufour95} with updated atomic data   
   \begin{figure}
    \centering
    \includegraphics[scale=0.45]{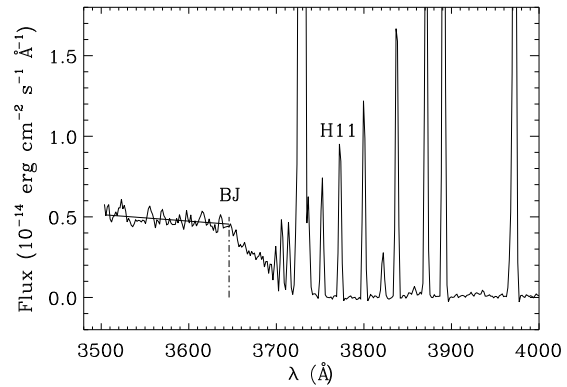}
    \caption{Section of a PMAS spectrum of the brightest spaxel at the apex (-5.5,-6.5). The fitted 
             continuum contribution to the right side of the BJ was substracted to the whole range 
	     showed.}
    \label{bj}
   \end{figure}
  \citep[see][]{garciarojasetal05}. Following the same procedure as \cite{mesadelgadoetal08} for 
  the determination of the physical conditions, we have assumed an initial \te = 10000 K to derive 
  a first approximation of \nel, then we calculate \te([\ion{O}{iii}]) and \te([\ion{N}{ii}]), 
  and iterate until convergence. We have not corrected the observed intensity of [\ion{N}{ii}] 
  5755 \AA\ for contribution by recombination when determining \te([\ion{N}{ii}]) because this 
  contribution is expected to be rather small in \hii\ regions with high ionization degree such as 
  the Orion Nebula \citep[e.g.][]{estebanetal04}. Bidimensional maps of \nel, \te([\ion{O}{iii}]), 
  \te([\ion{N}{ii}])/\te([\ion{O}{iii}]) ratio and \te([\ion{N}{ii}]) are shown in Figures 
  \ref{condfis}a, \ref{condfis}b, \ref{condfis}c, and \ref{condfis}d, respectively.\\
  In Figure \ref{condfis}a we can see that the highest densities are just reached at the positions 
  of \hh-N and \hh-S with values of about 10000 \cmc. Similar values were obtained by 
  \citet{mesadelgadoetal08} for \hh-S. For the gas outside the HH object, we find a \nel\ of about 
  4000 \cmc\ and the lowest densities ($\sim$2500 \cmc) at the low surface brightness region at the 
  northwest corner of the FOV. The typical errors in our density determination are between 200 and 
  250 \cmc.\\
  Figure \ref{condfis}b shows that \te([\ion{O}{iii}]) is rather constant in the whole FOV. 
  However, the highest values of \te([\ion{O}{iii}]) trace the form of \hh\ --with an almost 
  constant temperature of 8500 K-- reaching the peak value at the \hh-S position where shock 
  heating should be maximum. On the other hand, \te([\ion{N}{ii}]) shows a much wider range of 
  values across the FOV, but its spatial distribution is completely different to that of  
  \te([\ion{O}{iii}]). The maximum values of \te([\ion{N}{ii}]) are found just behind \hh-S or 
  ahead of the northwest border of the bow shock of \hh-N (see Figure \ref{condfis}d). The lowest 
  values are found in the inner part of \hh\ ($\sim$9500 K). In Figure~\ref{condfis}c, the 
  \te([\ion{N}{ii}])/\te([\ion{O}{iii}]) ratio illustrates that \te([\ion{N}{ii}]) is always higher 
  than \te([\ion{O}{iii}]) but the ratio tends to be closer to one in the HH object and especially 
  at \hh-N where the density is higher. The uncertainties of the temperatures are about 600 K for 
  \te([\ion{N}{ii}]) and 250 K for \te([\ion{O}{iii}]).\\ 
  The Balmer continuum temperature, \te(Bac), of the ionized gas has also been determined for most 
  of the individual spaxels of the FOV; this is the first bidimensional \te(Bac) map ever obtained 
  for an ionized nebula. This temperature depends on the ratio between the Balmer jump (BJ) and the 
  intensity of a given Balmer line. We have measured the BJ subtracting the continuum fitted at 
  both sides of the Balmer discontinuity at 3646 \AA\ (see Figure~\ref{bj}). The continuum fitted 
  at both sides of the BJ is defined over spectral ranges free of emission lines covering the 
  possible maximum baseline. The errors associated with the continuum subtraction are included in 
  the discontinuity measurements. Finally, we have computed \te(Bac) in K from the ratio of the BJ 
  to the H11 flux using the relation proposed by \cite{liuetal01}:
   \begin{figure}
    \centering
    \includegraphics[scale=0.68]{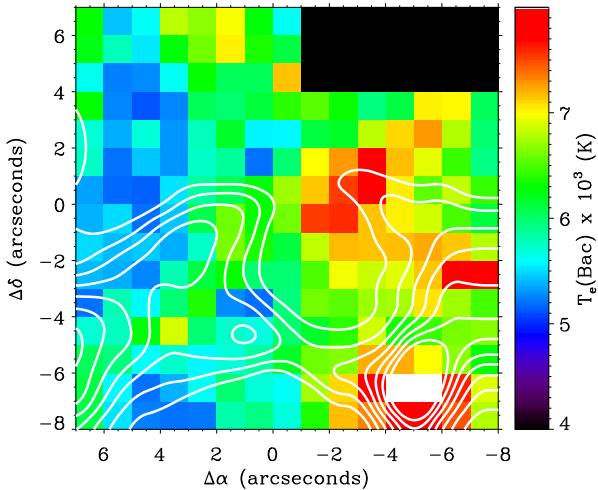} 
    \caption{Balmer temperature map with H$\alpha$ contours overplotted. The black rectangle on the 
             northwest corner corresponds to an area masked due to the bad determination of 
	     \te(Bac).}
    \label{tbalmer}
   \end{figure}	
  \begin{equation}
    { T_e({\rm Bac}) = 368\times\big( 1+ 0.259y^+ + 3.409y^{2+}\big)%
                              \Big(\frac{{\rm BJ}}{{\rm H11}}\Big)^{-\frac{3}{2}}}\,
  \end{equation} 
  where $y^+$ and $y^{++}$ correspond to the He$^+$/H$^+$ and He$^{++}$/H$^+$ ratios, respectively. 
  We have assumed for $y^+$ the value derived by \cite{estebanetal04} for their slit position at 
  the center of the Orion Nebula and $y^{++}$ = 0 due to the lack of \ion{He}{ii} lines in our 
  spectra. The effect of the interstellar extinction was taken into account in the calculations.\\  
  The \te(Bac) map is shown in Figure~\ref{tbalmer}. The \te(Bac) error was calculated by error 
  propagation in the equation (1), and it amounts to 22\% on average ($i.e.$, 1100 to 1900 K). We 
  can see that the lowest temperatures ($\sim$5500 K) are reached at the east half of the FOV, 
  while the west half shows values of about 7000 K, encompassing \hh-S and \hh-N. From the 
  comparison of Figure~\ref{tbalmer} and the rest of the maps, we can see that the spatial 
  distribution of \te(Bac) is something similar that of \te([\ion{O}{iii}]): both indicators show 
  the maximum values at \hh-S, \hh-N and the arc connecting both features.         
 \subsection{Chemical Abundances} \label{abpmas} 
   \begin{figure*}
    \centering
    \includegraphics[scale=0.6]{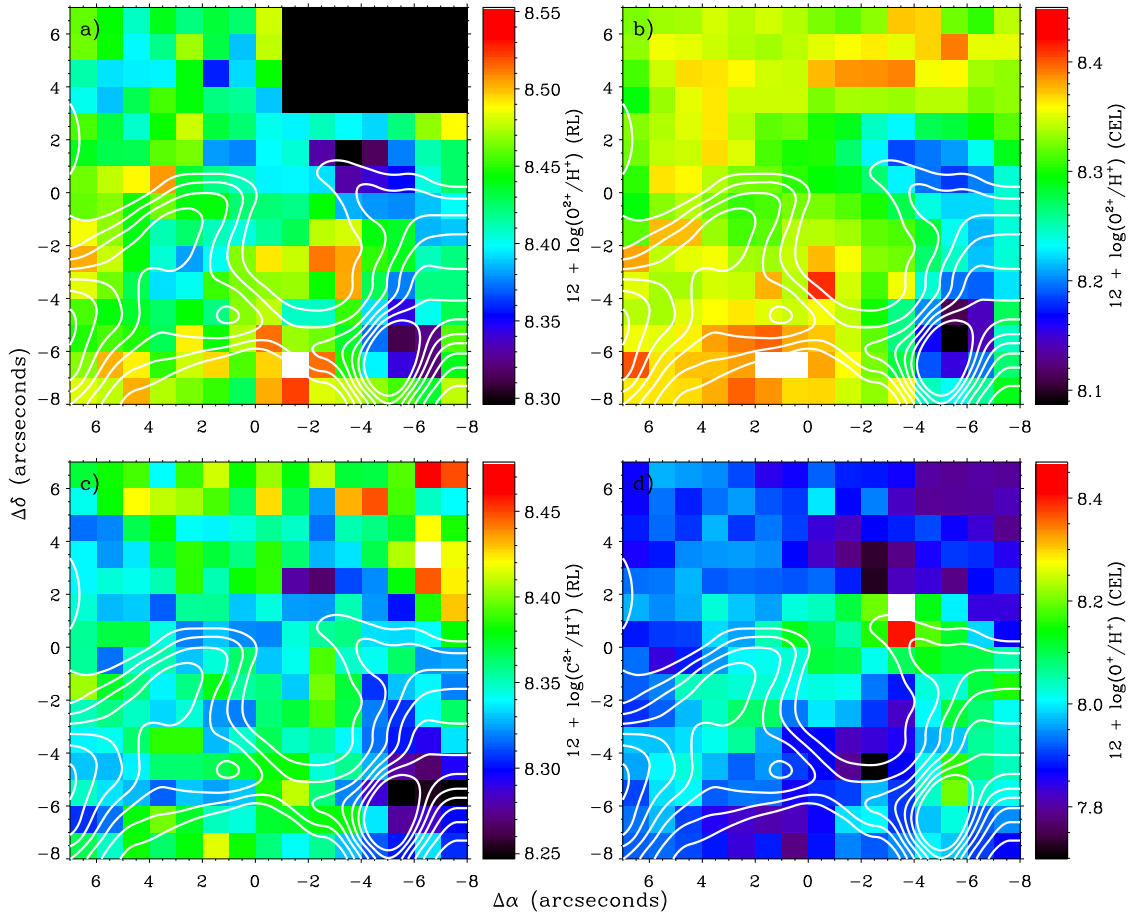} 
    \caption{Ionic abundance maps with H$\alpha$ contours overplotted: a) 12 $+$ 
             log(\ioni{O}{++}/\ioni{H}{+}) from RLs --the black rectangle on the northwest corner 
	     corresponds to an area masked due to the bad determination of the abundance, b) 12 $+$ 
	     log(\ioni{O}{++}/\ioni{H}{+}) from CELs, c) 12 $+$ log(\ioni{C}{++}/\ioni{H}{+}) from 
	     RLs and d) 12 $+$ log(\ioni{O}{+}/\ioni{H}{+}) from CELs.} 
    \label{abun}
   \end{figure*}
  The {\sc iraf} package {\sc nebular} has been used to derive ionic abundances of \ioni{N}{+}, 
  \ioni{O}{+} and \ioni{O}{2+} from the intensity of CELs. We have assumed no temperature 
  fluctuations in the ionized gas (\tf = 0) and a two-zone scheme: \te([\ion{N}{ii}]) for 
  \ioni{N}{+} and \ioni{O}{+} --the low ionization potential ions-- and \te([\ion{O}{iii}]) for 
  \ioni{O}{2+}. The errors in the ionic abundances have been calculated as a quadratic sum of the 
  independent contributions of errors in flux, \nel\ and \te. The spatial distribution of the 
  \ioni{O}{2+} and \ioni{O}{+} abundances are presented in Figures~\ref{abun}b and \ref{abun}d, 
  respectively. In these figures, we can see that the spatial distributions of both abundances are 
  completely different. The \ioni{O}{+}/\ioni{H}{+} ratio reaches the highest values just on the 
  bow shock, whereas \ioni{O}{2+}/\ioni{H}{+} shows an inverse behaviour, reflecting the different 
  ionization structure of the bow shock and the bulk of the background nebular gas. The higher 
  \ioni{O}{+}/\ioni{O}{2+} ratio at the bow shock is probably produced by the higher densities in 
  the shock gas, which increase the recombination rate of \ioni{O}{2+}. The abundace map of 
  \ioni{N}{+} --not presented here-- shows a very similar distribution to the \ioni{O}{+} map.\\ 
  The high signal-to-noise ratio of our spectra has permitted to detect and measure the weak RLs of 
  \ion{O}{ii} and \ion{C}{ii} in all the spaxels of the PMAS FOV. These RLs have the advantage that 
  their relative intensity with respect to a Balmer line is almost independent on \te\ and \nel, 
  largely unaffected by the possible presence of temperature fluctuations. In Figure~\ref{espmas} 
  we show a section of the spectra at spaxel (-5.5,~-6.5) around the \ion{O}{ii} lines. \\ 
  We have measured the blend of the \ion{O}{ii} 4649 and 4651 \AA\ lines --the brightest individual 
  lines of the \ion{O}{ii} multiplet 1-- in most of the spaxels of the FOV. In order to determine 
  the \ioni{O}{2+} abundance from this blend, we have combined equations (1) and (4) of 
  \cite{apeimbertpeimbert05} estimating the expected flux of all the lines of \ion{O}{ii} multiplet 
  1 flux with respect to H$\beta$, $I$(M1 \ion{O}{ii})/$I$(H$\beta$). These relations were used in 
  order to correct for the NLTE effects in the relative intensity of the individual lines of the 
  multiplet, though these are rather small in the Orion Nebula due to its relatively large \nel. 
  Then, the \ioni{O}{2+}/\ioni{H}{+} ratio is calculated by:
  \begin{equation}
    {\rm \frac{O^{2+}}{H^+}} = \frac{\lambda_{{\rm M1}}}{4861} \times %
                              \frac{\alpha_{eff}(H\beta)}{\alpha_{eff}({\rm M1})} \times %
                              \frac{I({\rm M1~O~II})}{I(H\beta)} ,
  \end{equation}            
  where $\alpha_{eff}(H\beta)$ and $\alpha_{eff}(\rm M1)$ are the effective recombination 
  coefficients for H$\beta$ and for the \ion{O}{ii} multiplet 1, respectively, and 
  $\lambda_{\rm M1}$ = 4651.5 \AA\ the representative mean wavelength of the whole multiplet. We 
  can also obtain the \ioni{C}{2+} abundance using an equation analogous to (2) but using the 
  specific quantities for this ion. Both abundances have been calculated assuming 
  \te([\ion{O}{iii}]) and with the effective recombination coefficients available in the literature 
  (\citealt{storey94} for \ioni{O}{2+} assuming LS coupling and \citealt{daveyetal00} for 
  \ioni{C}{2+}). Typical ionic abundance errors are 0.10 dex for \ioni{O}{2+} and 0.08 dex for 
  \ioni{C}{2+}. The ionic abundances obtained from RLs are shown in Figures \ref{abun}a and 
  \ref{abun}c.\\
  The spatial distributions of the \ioni{O}{2+} abundance obtained from CELs and RLs show that the 
  lower values are found at the positions of \hh-S and \hh-N and the arc connecting both features. 
  In contrast, the \ioni{O}{2+} abundance map obtained from CELs shows especially large values in 
  the zone inside the HH object at the bottom of the FOV. For a given spaxel, the 
  \ioni{O}{2+}/\ioni{H}{+} ratios obtained from RLs are always higher than those obtained from 
  CELs, with a typical difference of about 0.10 dex. Finally, the \ioni{C}{2+} abundance map shows 
  rather constant values of about 8.35 dex, with its minimum value also found at \hh-S. The 
   \begin{figure}
    \centering
    \includegraphics[scale=0.7]{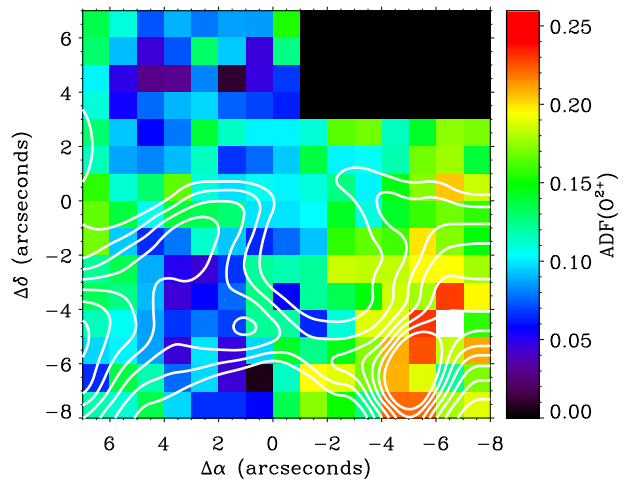} 
    \caption{\adfo\ map with H$\alpha$ contours overplotted. The black rectangle on the northwest 
             corner corresponds to an area masked due to the bad determination of the 
	     \ioni{O}{2+} abundance from RLs.}
    \label{adf}
   \end{figure}	
  qualitative behaviour of this map is rather similar to the \ioni{O}{2+} abundance spatial 
  distribution.\\ 
  For each spaxel, the \adfo\ has been calculated from the difference between the \ioni{O}{2+} 
  abundances obtained from RLs and CELs. In Figure~\ref{adf}, we show the spatial distribution of 
  the \adfo, which shows the highest values at and around \hh-S (up to 0.23 dex), the rest of the 
  spaxels show values between 0.05 and 0.20 dex with a typical error value of 0.12 dex. Excluding 
  the masked area showed in Figure~\ref{adf}, we have calculated an average value of 0.13 dex with 
  a standard deviation of 0.05 dex. \cite{mesadelgadoetal08} also found a peak of the \adfo\ of 
  about 0.30 dex at the same zone of \hh. Both values of the discrepancy are consistent within the 
  uncertainties, but the value obtained by those authors is, in principle, more reliable because of 
  the better signal-to-noise ratio of their spectra.      
\section{Discussion} \label{discu}
 \subsection{Temperature fluctuations} 
   \begin{figure*}
    \centering
    \includegraphics[scale=0.9]{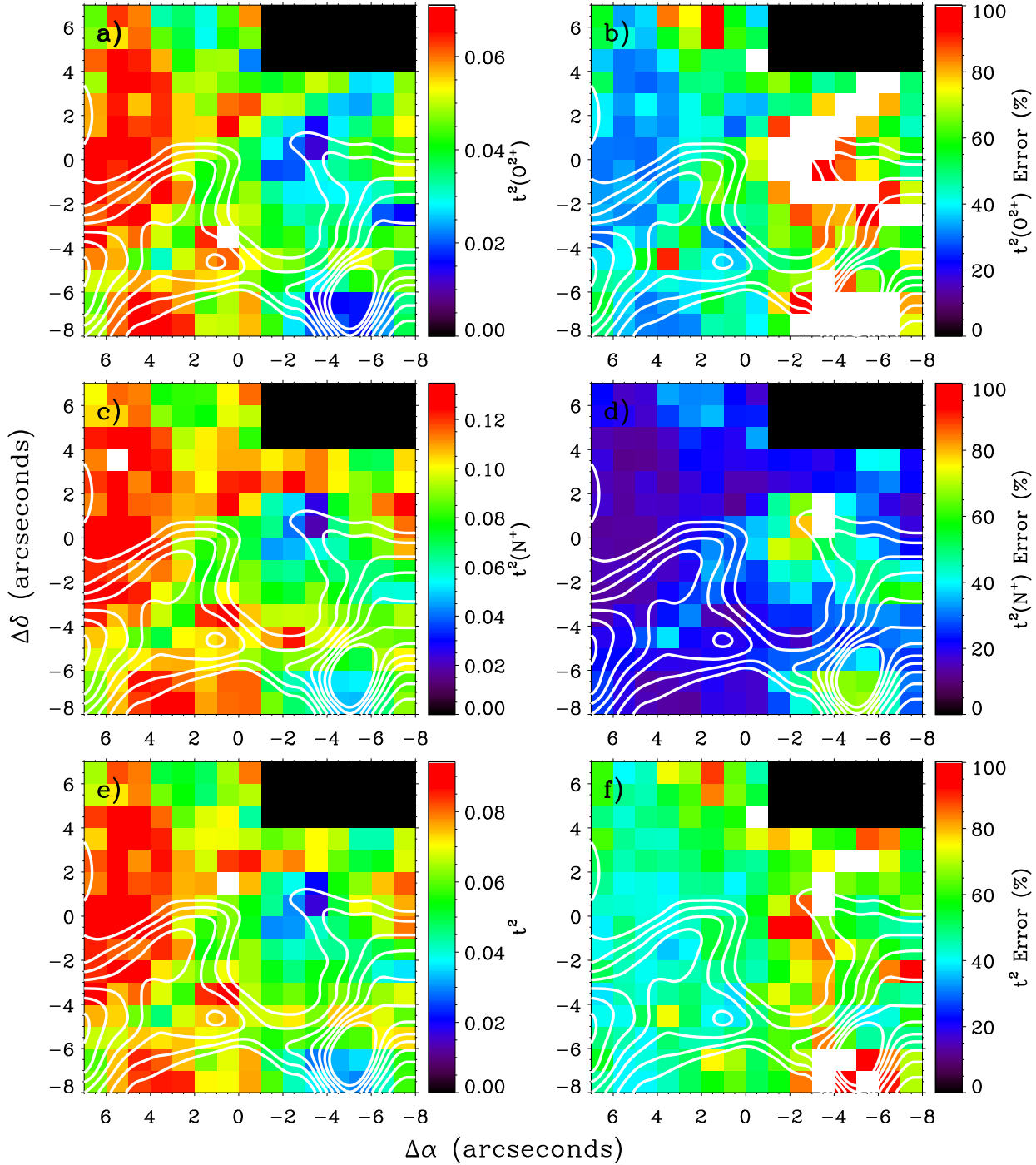} 
    \caption{\tf\ parameters and their respective error maps in percentage with H$\alpha$ contours 
             overplotted: a) and b) \tf(\ioni{O}{2+}) and its error, c) and d) \tf(\ioni{N}{+}) 
	     and its error, e) and f) \tf\ and its error. The white areas are where the error 
             is higher than the \tf\ value. The black rectangle on the northwest corner 
	     corresponds to an area masked due to the bad determination of \te(Bac).}
    \label{t2}
   \end{figure*}
  One of the possible explanations for the AD problem in ionized nebulae is provided by the 
  presence of temperature fluctuations \citep[e.g.][]{garciarojasesteban07}. Following 
  \cite{peimbert67}, the temperature fluctuation over the observed volume of a nebula can be 
  parameterized by
  \begin{equation}
    { T_0 = \frac{\int{T_e n_e n_i dV}}{\int{n_e n_i dV} } }\ ,
  \end{equation}
  where $T_0$ is the average temperature of the nebula and $n_i$ is the ion density, and 
  \begin{equation}
    { t^2 = \frac{\int{(T_e - T_0)^2 n_e n_i dV}}{T_0^2\int{n_e n_i dV}}} \ ,
  \end{equation}    
  where \tf\ is the mean-square electron temperature fluctuation.\\
  There are several methods to obtain \tf\ and $T_0$ along the line of sight \citep[see][]%
  {peimbertetal04}. One possibility is to compare the \te\ obtained from independent methods. 
  Following \cite{peimbert67}, the temperature derived from the BJ depends on \tf\ and $T_0$ as:
  \begin{equation}
   { T_e(Bac) = T_0 ( 1 - 1.70t^2)}\ ,
  \end{equation}
  and the temperatures derived from the ratio of CELs can be written as \citep{peimbertcostero69} 
  \begin{equation}
   {  T_e({\rm h}) = T_0({\rm h}) \Big[ 1 + %
                         \frac{1}{2}\big(\frac{91300}{T_0({\rm h})} -3\big)t^2({\rm h}) \Big] }\ ,
  \end{equation} 
  and
  \begin{equation}
   {  T_e({\rm l}) = T_0({\rm l}) \Big[ 1 + %
                        \frac{1}{2}\big(\frac{69000}{T_0({\rm l})} -3\big)t^2({\rm l}) \Big] }\ ,
  \end{equation}
  where \te(l) and \te(h) are the electron temperatures for the low --\te([\ion{N}{ii}])-- and 
  high --\te([\ion{O}{iii}])-- ionization zone, respectively. On one hand, we can obtain the 
  \tf(\ioni{O}{2+}) and $T_0$(\ioni{O}{2+}) combining equations (5) and (6). On the other 
  hand, \tf(\ioni{N}{+}) and $T_0$(\ioni{N}{+}) can be obtained combining expressions (5) 
  and (7). To obtain the mean square temperature fluctuation for the entire volume, \tf, we 
  have weigthed the relative importance of \tf(\ioni{N}{+}) and \tf(\ioni{O}{2+}) in the observed 
  volume by using equation 16 of \cite{apeimbertetal02}, assuming that in this region \ioni{N}{+} 
  and \ioni{O}{+} coexist in the same volume. In Figures~\ref{t2}a, \ref{t2}c and \ref{t2}e, the 
  \tf(\ioni{O}{2+}), \tf(\ioni{N}{+}) and $<t^2>$ maps are presented. Their associated error maps 
  are also shown in Figures \ref{t2}b, \ref{t2}d and \ref{t2}f.\\
  The inspection of Figures~\ref{t2}b, \ref{t2}c and \ref{t2}e indicates that the spatial 
  distribution of \tf\ values are rather similar, showing somewhat lower \tf\ values at the 
  positions of \hh-S and \hh-N. We have calculated the mean value of $<t^2>$ in the area where this 
  quantity has been determined weighting the individual values of the fluctuation for each spaxel 
  by its associated error, obtaining $<t^2>$ $\approx$ 0.061$\pm$0.022.\\
  Another possibility to obtain the \tf\ parameter is assuming that the \adfo\ is produced by the 
  presence of temperature fluctuations. Following this assumption, we have used equations (9), 
  (10) and (11) from \cite{peimbertetal04} to derive the associated \tf(\ioni{O}{2+}) for each 
  spaxel that --as expected-- shows a spatial distribution analogous to the ADF map shown in 
  Figure \ref{adf}. In this case, the mean value of \tf\ is only slightly lower than that obtained 
  from the data of Figure \ref{t2}a. A remarkable result is that the \tf\ maps obtained from the 
  two different methods --comparison of temperatures and from the \adfo-- are qualitatively 
  different suggesting, in this case, that there is not a genetic relationship between \tf\ and 
  \adfo.\\
  Finally, we have determined the \tf\ in the plane of the sky, $t^2_A$, using equations (11) and 
  (12) of \cite{mesadelgadoetal08}, averaging the bidimensional point-to-point variations of 
  \te(\ioni{O}{2+}) and \te(\ioni{N}{+}). The values obtained are $t^2_A$(\ioni{O}{2+}) 
  $\sim$ 0.0004 and $t^2_A$(\ioni{N}{+}) $\sim$ 0.0023, in agreement with the long-slit results by 
  \cite{mesadelgadoetal08}, and substantially lower than the \tf\ values obtained with the methods 
  described in the previous paragraphs, which correspond to the fluctuations along the line of 
  sight.     
 \subsection{Correlations among the ADF and other nebular properties} \label{correl}
   \begin{figure*}
    \centering
    \includegraphics[scale=0.55]{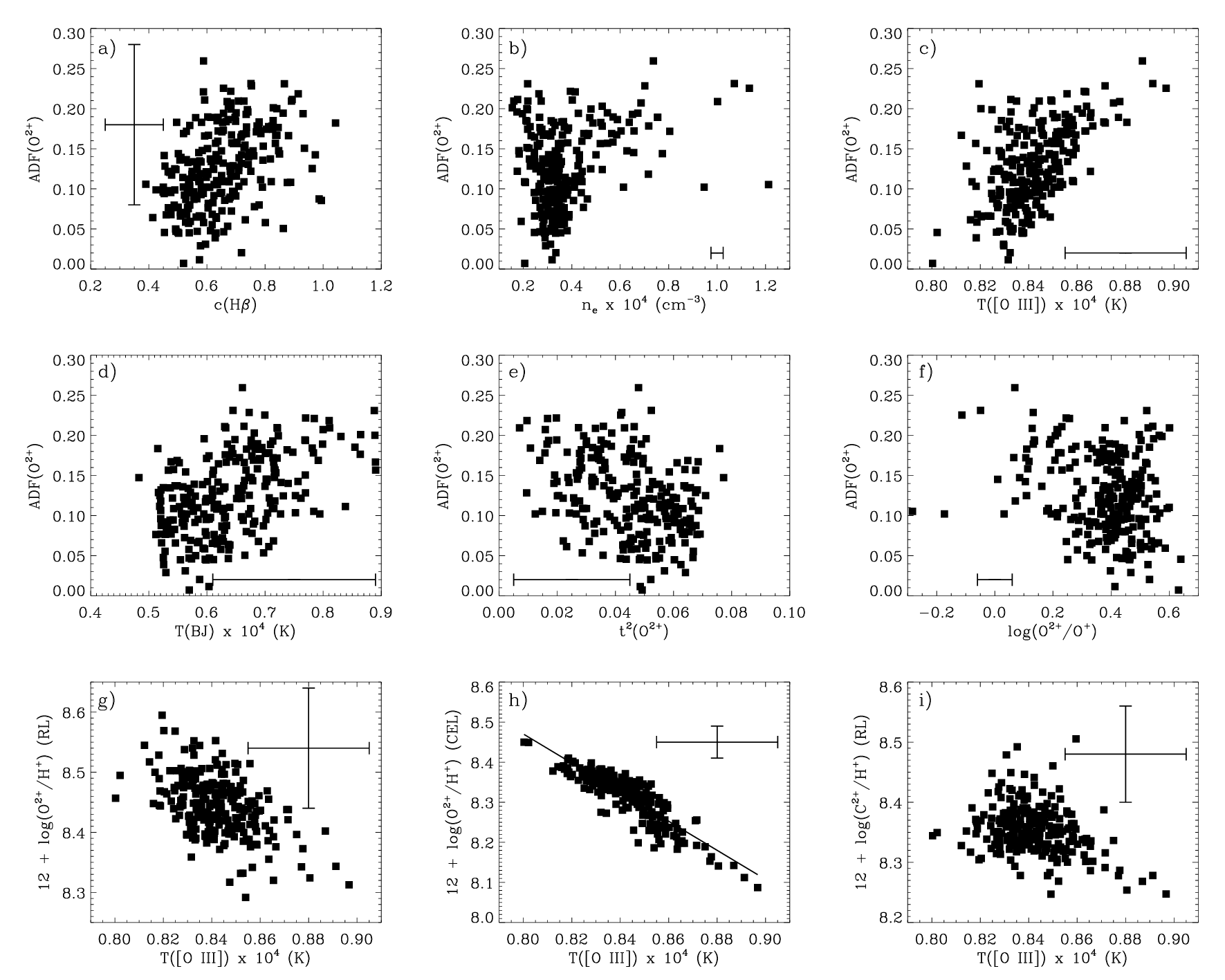} 
    \caption{Correlations among different nebular properties obtained from PMAS maps. From a) to f) 
             the \adfo\ is represented with respect to c(H$\beta$), \nel, \te([\ion{O}{iii}]), 
	     \te(BJ), \tf[\ion{O}{iii}] and log(\ioni{O}{2+}/\ioni{O}{+}). The mean error bar in 
	     the \adfo\ is only shown in a). From g) to i) the logarithmic abundances of 
	     \ioni{O}{2+} from RLs, \ioni{O}{2+} from CELs and \ioni{C}{2+} from RLs are 
	     represented $vs.$ \te([\ion{O}{iii}]).} 
    \label{adfcor}
   \end{figure*}
  In this section, we investigate some relationships among physical properties derived in the \hh\ 
  PMAS field. In Figure~\ref{adfcor} the possible correlations among the \adfo, \te, \nel, 
  c(H$\beta$), \tf\ and ionic abundances determined from CELs and RLs are presented.\\
  The dependence of the \adfo\ with respect to c(H$\beta$), \nel, \te([\ion{O}{iii}]), \te(BJ), 
  \tf(\ioni{O}{2+}) and \ioni{O}{2+}/\ioni{O}{+} ratio is shown from Figures~\ref{adfcor}a to 
  \ref{adfcor}f. Figure~\ref{adfcor}a --\adfo\ $vs.$ c(H$\beta$)-- does not show a clear trend, 
  indicating that the ADF does not depend on the amount of dust present in the line of sight of 
  each spaxel of the PMAS field. As \cite{mesadelgadoetal08} also obtained from their long-slit 
  data, there seems to be no correlation between the ADF and \nel\ (see Figure \ref{adfcor}b) as 
  well as between the ADF and the ionization degree of the gas (see Figure~\ref{adfcor}f).\\ 
  In Figures~\ref{adfcor}c and \ref{adfcor}d, we can see that positive correlations are obtained 
  between \adfo\ $vs.$ \te([\ion{O}{iii}]) and \adfo\ $vs.$ \te(BJ), respectively. This is expected 
  because these three quantities show larger values at \hh-S. As we can see in 
  Figures~\ref{adfcor}c and \ref{adfcor}d, these correlations do not seem very reliable considering 
  the very narrow baseline of \te([\ion{O}{iii}]) and \te(BJ) covered by the data and the large 
  error bars with respect to the baseline.\\
  A rather unclear negative correlation between the \adfo\ and the fluctuation parameter 
  \tf(\ioni{O}{2+}) along the line of sight can be guessed in Figure~\ref{adfcor}e. However, the 
  large error bars in both quantities do not permit to establish any correlation. This result 
  suggests the ADF and \tf\ may be independent.\\
  In Figures~\ref{adfcor}g, \ref{adfcor}h and \ref{adfcor}i, we represent the \ioni{O}{2+} 
  abundances from RLs, \ioni{O}{2+} abundances from CELs and \ioni{C}{2+} abundances from RLs 
  $vs.$ \te([\ion{O}{iii}]), respectively. In the case of the abundances from RLs, we can 
  observe negative correlations with rather low Spearman correlation coefficients ($\rho$ = $-$0.42 
  and $\rho$ = $-$0.28 for Figures~\ref{adfcor}g and \ref{adfcor}i, respectively). This apparent 
  trend is mainly produced by the behaviour observed at \hh-S and \hh-N, which are precisely the 
  zones where the \ioni{O}{2+} and \ioni{C}{2+} abundances obtained from RLs are lower and 
  \te([\ion{O}{iii}]) is higher. From integral field spectroscopy, \cite{tsamisetal08} have found 
  rather clear positive correlations between ionic abundances obtained from RLs and \te\ in three 
  PNe, exactly the opposite trend we observe in this FOV of the Orion Nebula. These authors argue 
  that their results provide evidence for the existence of two distinct components of highly 
  ionized gas at very different temperatures. However, in this case, we observe the expected 
  natural behaviour of an ionized nebula --the \ioni{O}{2+} abundance from RLs increases when 
  the temperature decreases-- due to the fact that \ioni{O}{2+} is a dominant coolant in the gas 
  phase and the presence of spatial changes of the \ioni{O}{2+}/\ioni{O}{+} ratio across the FOV. 
  Therefore, the presence of metal-rich droplets does not seem to be supported by our results. On 
  the other hand, the \ioni{O}{2+}/\ioni{H}{+} ratio from CELs shows a much tighter correlation 
  with \te([\ion{O}{iii}]) due to the lower uncertainties of these abundances.  We have fitted the 
  following linear relation:
  \begin{equation}
    {12+{\rm log(\frac{O^{2+}}{H^+}})= (11.4\pm0.1)+(-3.6\pm0.1)T_4}
  \end{equation}  
  where $T_4$ is \te([\ion{O}{iii}]) in units of 10$^4$ K and we have obtained a linear correlation 
  coefficient $r$ = $-$0.90. 
\section{Conclusions} \label{conclu}
 In this paper, we present results from integral field spectroscopy of an area of 
 15$\arcsec\times$15$\arcsec$ covering the head of \hh\ in the Orion Nebula. The FOV comprises the 
 bright regions known as \hh-S and \hh-N. We have obtained maps of relevant emission line ratios, 
 physical conditions and ionic abundances, including \ioni{O}{2+} and \ioni{C}{2+} abundances 
 determined from recombination lines (RLs). Additionally, we have obtained maps of other 
 interesting nebular properties, as the temperature fluctuation parameter, \tf, and the abundance 
 discrepancy factor of \ioni{O}{2+}, \adfo, which is defined as the difference between the 
 \ioni{O}{2+}/\ioni{H}{+} ratios determined from RLs and collisionally excited lines (CELs).\\
 We find that the flux distributions of the [\ion{O}{iii}] and \ion{O}{ii} lines are rather 
 similar and that the HH object is comparatively much brighter in the lines of low-ionization 
 potential ionic species and especially in [\ion{Fe}{iii}]. The \nel\ is about 4000 \cmc\ in most 
 of the FOV and higher --about 10000 \cmc-- at \hh-S and \hh-N. The \te([\ion{O}{iii}]) map shows 
 a narrow range of variation, but the values are higher at \hh-S. On the other hand, 
 \te([\ion{N}{ii}]) shows larger variations and a very different spatial distribution, being 
 higher at the northern and eastern edges of \hh-N and \hh-S, respectively, a likely consequence 
 of the ionization stratification or the presence of some shock excitation in the knots of the HH 
 object. We have obtained --for the first time in an ionized nebula-- the \te(Bac) map, which 
 follows closely that of \te([\ion{O}{iii}]).\\ 
 The \ioni{O}{+}/\ioni{H}{+} ratio map reaches the highest values just on the arc that delineates 
 the north of \hh\ from the east and west edges of the FOV, whereas the map of \ioni{O}{2+} 
 abundance obtained from CELs shows an inverse behaviour, probably produced by the higher densities 
 in the shock gas that increase the recombination rate of \ioni{O}{2+}. The spatial distributions 
 of the \ioni{O}{2+} abundance obtained from CELs and RLs agree in showing the lower values at the
 positions of \hh-S and \hh-N and the arc connecting both features. However, for a given spaxel, 
 the \ioni{O}{2+}/\ioni{H}{+} ratios obtained from RLs are always about 0.10 dex higher than those 
 obtained from CELs. The map of the \adfo\ shows the highest values at and around \hh-S.\\ 
 We have determined --for the first time in an ionized nebula-- the \tf\ map of the FOV from the 
 comparison of \te(Bac) and the other \te\ values determined from collisionally excited line 
 ratios, finding that it does not match the \adfo\ map. This result supports that the AD and 
 temperature fluctuations are independent phenomena. \\
 Finally, we have found weak correlations between the \adfo\ and \te\ values of the different 
 spaxels of the FOV and between the \ioni{O}{2+}/\ioni{H}{+} ratios obtained from RLs and CELs and 
 \te([\ion{O}{iii}]). The negative correlation between the \ioni{O}{2+} abundances obtained from 
 RLs and \te([\ion{O}{iii}]) does not support the predictions of the chemical inhomogeneity models 
 of \cite{tsamispequignot05} and \cite{stasinskaetal07}. 
\section*{Acknowledgments}
We thank Sebasti\'an F. S\'anchez and the personnel of Calar Alto Observatory for their help with 
the observations. We are grateful to M. Peimbert, G. Stasi\'nska and W. Henney for several 
suggestions and discussions. We also thank the referee of the paper for his/her positive comments. 
This work has been funded by the Spanish Ministerio de Ciencia y Tecnolog\'\i a(MCyT) under project 
AYA2004-07466 and Ministerio de Educaci\'on y Ciencia (MEC) under project AYA2007-63030 and 64712.


\label{lastpage}

\end{document}